# ON THE PROBLEM OF THE STAGGERED FIELD IN CuGeO$_3$ DOPED WITH MAGNETIC IMPURITIES

S.V.Demishev

*Low Temperatures and Cryogenic Engineering Department, A.M.Prokhorov General Physics Institute of Russian Academy of Sciences, Vavilov street, 38, 119991 Moscow, Russia*
*E-mail: demis@lt.gpi.ru*

**Abstract**

The magnitude of the staggered field is calculated from the EPR data for CuGeO$_3$ doped with Co and Fe magnetic impurities. It is found that this parameter demonstrate an anomalous temperature and magnetic field dependence probably due to (i) the specific mechanism of the staggered field generation in doped CuGeO$_3$ and (ii) a competition between antiferromagnetic interchain exchange and staggered Zeeman energy.

Resent works of Oshikawa and co-workers have pointed out the importance of staggered field (SF) in genesis of the anomalous static and dynamic magnetic properties of S=1/2 one-dimensional (1D) antiferromagnetic (AF) spin chains [1-3]. In particular in electron paramagnetic resonance (EPR) the presence of the SF manifests itself via low temperature growth of the *g*-factor (or decrease of the resonant field $B_{res}$) and broadening of the line width *w* [1,2,4],

$$\Delta g(T) = \Delta g_{sf}(T) = 0.344057 \frac{J\mu_B^2 h^2}{(k_B T)^3} \ln\left(\frac{J}{k_B T}\right), \qquad (1)$$

$$w(T) = w_{sf}(T) = 0.68571 \frac{J\mu_B h^2}{(k_B T)^2} \ln\left(\frac{J}{k_B T}\right), \qquad (2)$$

where $\Delta g(T)$ denotes temperature dependent correction to the *g*-factor, *J* and *h* stand for the exchange integral and magnitude of the staggered field respectively. As long as certain quasi one-dimensional magnets really show behavior qualitatively corresponding to equations (1) and (2) several attempts have been made to compare theoretical predictions with experiment. Up to now the most detail study have been carried out in the case of Cu-benzoate (for review see Ref. 1,2,5,6), where in addition to the anomalous EPR described by (1) and (2) a transition to the regime of magnetic field induced gap and resonant modes of the soliton (breather) types have been observed in agreement with predictions [1-2]. Similar investigation have been performed by Ohta *et al.* for the low-dimensional magnet BaCu$_2$(Si$_{0.35}$Ge$_{0.65}$)$_2$O$_7$ [7,8]. In both cases the expressions for the resonant field shift and line width have been applied directly to analysis of the experiment and demonstrated a reasonable agreement with the measured data. However, in order to improve quality of the fitting procedure in [7,8] an unjustified in theory empirical terms have been added to the expression for the resonant field.

Later Demishev *et al.* have suggested that model of the staggered field can be used for the explanation of the EPR peculiarities in CuGeO$_3$ doped with magnetic impurities [4]. In contrast to [5-8] the analysis procedure used in [4] have been different from the aforementioned schema. It follows from equations (1) and (2) that $\Delta g_{sf}(T)$ and $w_{sf}$ are linked by the universal relation [4]

$$\frac{w_{sf}}{\Delta g_{sf}} = 1.99 \frac{k_B T}{\mu_B} \qquad (3)$$

independent of the exchange integral and staggered field values. Although equation (3) does not contain any free parameters it works surprisingly well for $CuGeO_3$ doped with Fe and Co impurities [4]. In spite that the magnitudes of the line width and $g$-factors shift for $CuGeO_3$:Fe and $CuGeO_3$:Co were quite different, the universal character of the temperature dependence of the ratio $w_{sf}/\Delta g_{sf}(T)$ was conserved thus providing an argument for the presence of the staggered field in doped $CuGeO_3$ [4].

It follows from the above consideration that there are reasons to trust in theoretical calculations [1-2] and they may be applied to analysis of the experiment in a different ways. For example, the equations (1) and (2) can be used for finding of the staggered field magnitude. Therefore the EPR experiments may provide direct information concerning the staggered component in the local magnetic field in the sample and its dependence on the external parameters (magnetic field, temperature *etc.*). As far as we know this aspect of the staggered field problem have not been touched up to now, and we will address it in the present short note taking doped $CuGeO_3$ as an experimental example.

Before starting of the suggested procedure it is worth to think why staggered field may appear in doped $CuGeO_3$. In theory staggered field is a magnetic field that changes direction alternatingly. In the considered works [1-3] the effects of the staggered field are described by the term in Heisenberg Hamiltonian

$$\hat{H}_{sf} = -h \cdot \sum_{i,j,k} (-1)^i S^x_{i,j,k}. \qquad (4)$$

In equation (4) the $S^x_{i,j,k}$ is the x-component of the spin ½ operator on numbered (i, j, k) site. As long as the external magnetic field $B$ is aligned along $z$-axis, the staggered field should lie in direction perpendicular to external magnetic field. In addition it is generally assumed that staggered field magnitude is proportional to the external magnetic field $h \sim B$ [1-8].

The mechanisms of the staggered field generation may be different in different experimental objects [3]. The most popular up to now where (i) staggered component of the $g$-tensor and (ii) staggered Dzyaloshinskii-Moriya (DM) interaction [1-8]. Both mechanisms require certain low symmetry of the crystal structure and thus can be hardly applicable to the case of the doped $CuGeO_3$. Indeed the EPR data for pure and doped with non-magnetic impurities $CuGeO_3$ provides no sign the anomalies described by equations (1) and (2) in the range $T > T_{SP}$, where spin chains are homogeneous and theory [1-2] is applicable (here $T_{SP}$ denotes temperature of the spin-Peierls transition). Instead the line width in this case is terminated by exchange anisotropy [2,4] and $w(T) \to 0$ when $T \to 0$. Although the doping of $CuGeO_3$ by 1% of Fe [4,9,10] or by 2% of Co [4,11,12] completely damps both spin-Peierls and Neel transitions at least up to 0.35 K [13] thus providing homogeneity of the AF spin chains it is difficult to assume that doping on this level can change crystal symmetry and exchange integral in chains. Therefore in our opinion the mechanisms of the generation of the staggered field in $CuGeO_3$ based on staggered $g$-tensor or DM interaction appear to be unlikely.

An alternative idea concerning formation of the staggered field has been considered in [3]. If the low-dimensional magnetic system consists of two sublattices and possesses a strong intralattice coupling and a weak interlattice coupling, than assuming that one of sublattices is ordered antiferromagnetically we see that it will induce staggered field on the other sublattice. When applying this approach to the real "1D magnets" it is reasonable to suggest that interchain exchange, which makes these solids just quasi one-dimensional from one hand, will open an opportunity for generation of the staggered field in the sample. Coming back to the case of doped $CuGeO_3$ it is necessary to remind that in this solid the interchain interaction is not weak, being about 10% of the intrachain interaction [14]. Our hypothesis is that the doping of magnetic impurities may *locally* alter the intrachain interaction and likely affect *local* magnetic order. As a result a spin clusters (nearly AF ordered) will be formed and these clusters will possess staggered field die to the aforementioned induction mechanism. Actually these clusters, where the correlation of spins is stronger than average, is a fingerprint of the Griffiths phase characteristic to various disordered spin systems [15-17]. In our previous publications [9-12] we have provided experimental arguments that transition into Griffiths phase in $CuGeO_3$ doped with Fe and Co occurs below 20-70 K, thus the new suggestion is that local formation of the staggered field at low temperatures and quantum critical regime, for which the Griffiths phase serves as a ground state, are closely connected.

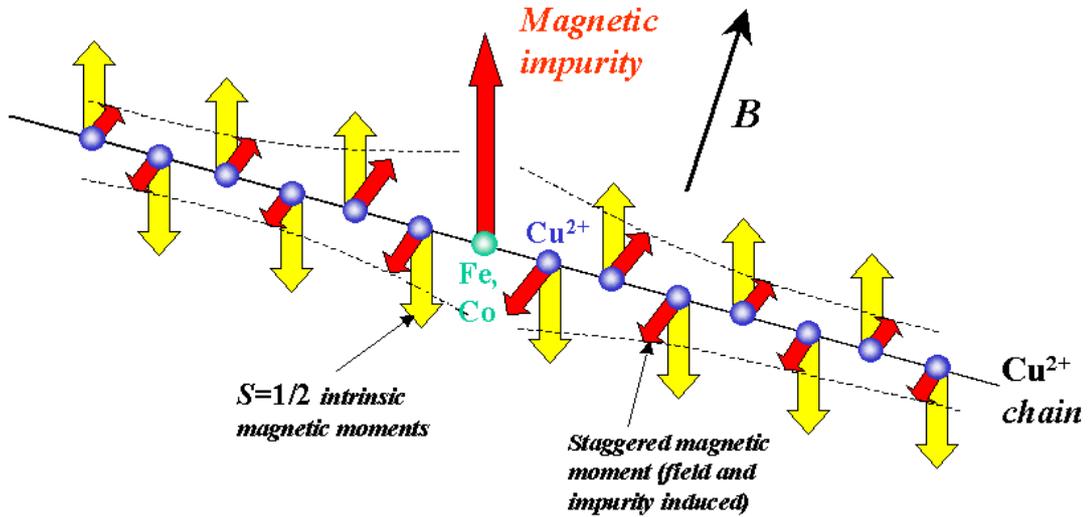

**Fig. 1.** Expected structure of the staggered magnetisation in $CuGeO_3$ doped with magnetic impurities (only one $Cu^{2+}$ chain is shown).

Following the above consideration we conclude that in $CuGeO_3$ the staggered field may be spatially inhomogeneous, may depend on temperature due to magnetic impurities and interchain interactions, and, finally the simple linear relation between external magnetic field and staggered field, $h \sim B$, may be violated. The latter possibility is different from the case, when the staggered field is generated by staggered *g*-tensor or DM interaction where the condition $h \sim B$ should always hold, and is explained by contribution to the local magnetic fields in the sample from the magnetic moment of magnetic impurity, so that $h(B=0) \neq 0$. The expected qualitative picture of the staggered field in doped $CuGeO_3$ is presented in fig. 1.

As long as in or case $h = h(\vec{r}, T, B)$ the equations (1) and (2) should be somehow averaged over spatial coordinate before applying to the considered case. Moreover, the extra temperature dependence of the staggered field may change the theoretical asymptotics $w \sim 1/T^2$ and $\Delta g \sim 1/T^3$ (equations (1) and (2)). Nevertheless the universal relation (3) may be less affected due to the fact that all averaged quantities are coming into equations (1) and (2) in a same way, and will work well for $CuGeO_3$ doped with magnetic impurities [4].

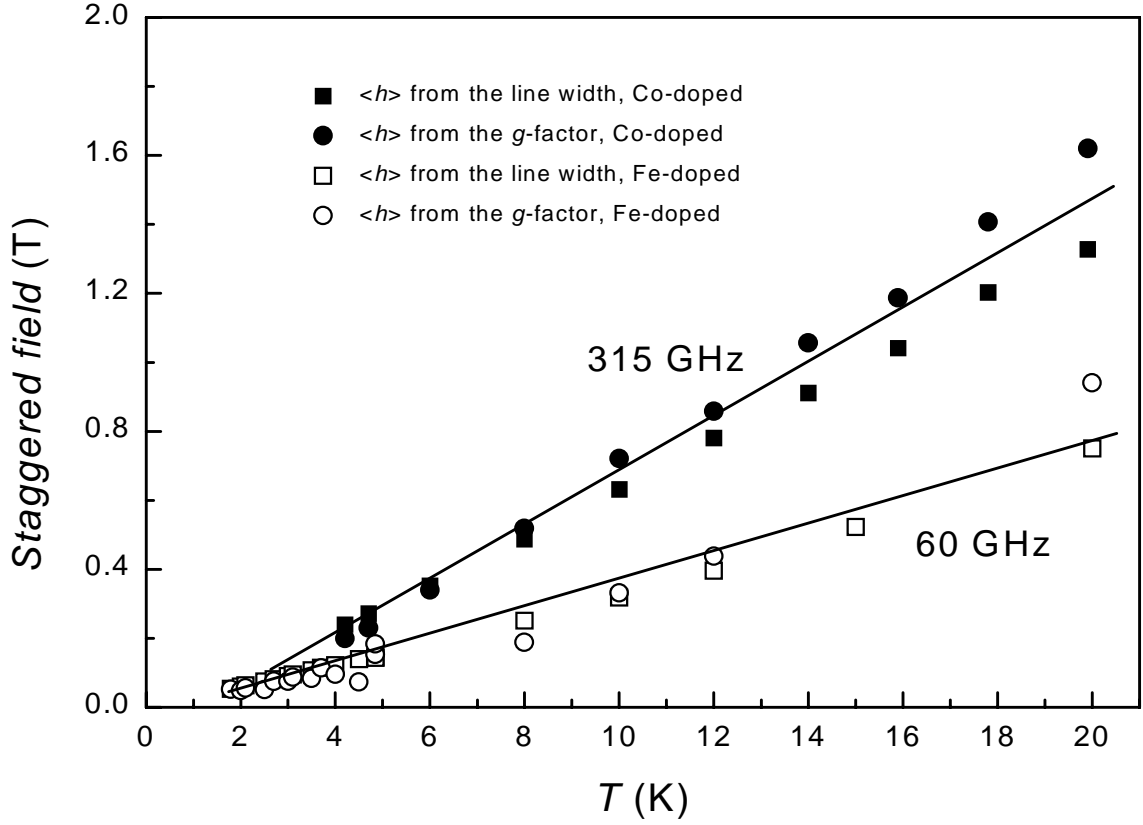

Fig. 2. Temperature dependence of the staggered field in $CuGeO_3$:Co and $CuGeO_3$:Fe from EPR measurements. Solid lines are guides for the eyes.

Now we are ready to use equations (1) and (2) as definitions of the averaged staggered field $<h>$ and apply them to experimental data reported in Ref. 4,9-12 for $CuGeO_3$ doped with 1% of Fe or 2% of Co. The procedure of the determination of $w_{sf}$ is given in [4] and, following [4,9-12], it is supposed that $\Delta g = g(T) - g(30\ K)$. The results are presented in fig. 2 (temperature dependence) and fig. 3 (field dependence). In calculations the value $J = 120\ K$ [14] have been used. First of all it is necessary to mark the good coincidence between staggered field magnitudes found from the line width and $g$-factor (fig. 2), that means the validity of the universal relation (3). However the obtained quantity $<h>$ demonstrates strong temperature dependence: the magnitude of SF decreases 7-8 times when temperature is lowered from 20 K to 4 K. At the same time following the initial consideration [1-2] for the cases of staggered $g$-tensor or DM interaction one can expect $h(T) = const$.

The field dependence of the averaged SF $<h>$ also exhibits considerable deviations from the theoretical suppositions [1-3]. Indeed instead of relation $h \sim B$ suggested in [1-3] the experimental data in fig. 3 are better represented by

$$<h(B)> = h(0) + A \cdot B, \qquad (5)$$

i.e. $\langle h(B=0)\rangle \neq 0$. Therefore in addition to the external magnetic field induced component, which is proportional to $B$, an intrinsic SF appears in the sample. In our opinion this effect is specific to the case of $CuGeO_3$ doped with magnetic impurities as it was qualitatively described above. It is worth to note, that best fits of the experimental data with equation (5) have provided very close values of the coefficient $A$, namely 0.0020 for Fe-doped case and 0.0024 for Co-doped case. Consequently the "staggered field susceptibility", which define the part of the staggered field dependent on the external magnetic field, appears to be weakly dependent on the impurity studied. At the same time data in fig. 3 provides values $\langle h(0)\rangle = 0.052$ T and $\langle h(0)\rangle = 0.077$ T for Fe and Co impurities respectively. This discrepancy can be easily explained by the fact that impurity concentration in $CuGeO_3$:Co was two times higher than in $CuGeO_3$:Fe and hence the contribution from the impurity magnetic moment should be bigger in the Co-doped case.

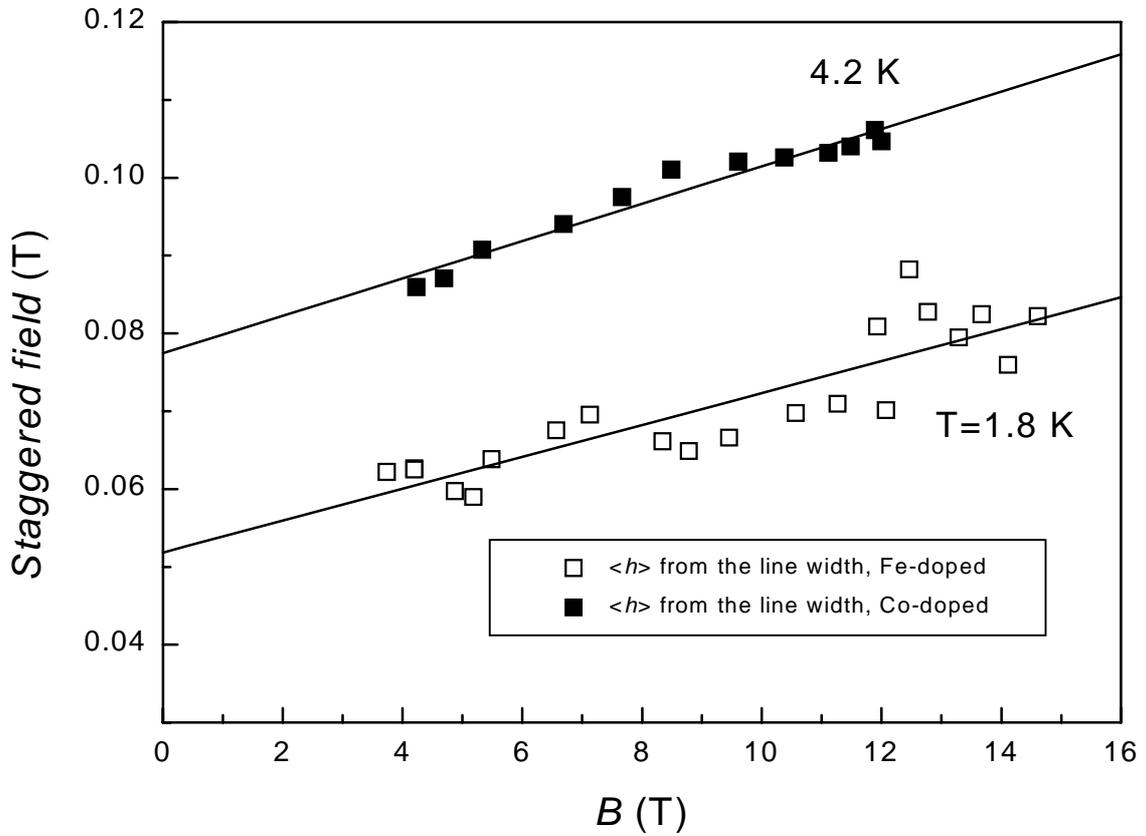

Fig. 3. Field dependences of the staggered field in doped $CuGeO_3$. Solid lines represent best fits with equation (5).

Interesting that experimental data on antiferromagnetic resonance in Ni and Zn doped $CuGeO_3$ reported by Smirnov et al. [18, 19] suggests a characteristic gap frequency about 20 GHz and spin-flop field about 1 T. Therefore the magnetic fields characteristic to AF case, namely $H_E$ and $H_A$ [20], could be comparable with the magnitudes of the SF $\langle h \rangle$ reported in the present work (fig. 2,3). In our opinion this favors the idea about competition between interchain antiferromagnetic interactions and staggered Zeeman energy [3], which is characteristic to aforementioned induction mechanism of the staggered field formation. It is possible that this competition may give rise to the temperature dependence of $\langle h \rangle$ shown in fig. 2 as long as in doped $CuGeO_3$ antiferromagnetic correlations becomes stronger at low temperatures and thus may decrease the magnitude of the SF. The checking of the latter supposition requires further theoretical studies.

In conclusion we wish to point out that analysis performed in the present work indicates that at least in the certain cases the SF behavior may be more complicated than thought initially [1-3] and magnitude of the staggered field may acquire specific field and temperature dependence. As a practical consequence of this fact will be deviations of the EPR line width and *g*-factor from the theoretical predictions (equations (1) and (2)) whereas the universal relation (3) will still work. Checking of this supposition for various experimental systems and further theoretical studies in this direction could be rewarding.

I am grateful to Professor H.Ohta, Dr. S.Okubo, Dr. A.V.Semeno and Dr. N.E.Sluchanko for helpful discussions. The financial support from the Russian Science Support Foundation, Russian Foundation for the Basic Research grant 04-02-16574 and programme of the Russian Academy of Sciences "Strongly Correlated Electrons" is acknowledged.